# What are the microscopic events of colloidal membrane fouling?

J. Lohaus[a], Y.M. Perez[a], M. Wessling[a,b,*]

[a]*RWTH Aachen University, Chemical Process Engineering, Turmstrasse 46, D-52064 Aachen, Germany*
[b]*DWI-Leibniz-Institute for Interactive Materials, RWTH Aachen University, Forckenbeckstrasse 50, D-52074 Aachen, Germany*

## Abstract

Due to the complex interplay between surface adsorption and hydrodynamic interactions, representative microsocpic mechanisms of colloidal membrane fouling are still not well understood. Numerical simulations overcome experimental limitations such as the temporal and spatial resolution of microscopic events during colloidal membrane fouling: they help to gain deeper insight into fouling processes. This study uses coupled computational fluid dynamics - discrete element methods (CFD-DEM) simulations to examine mechanisms of colloidal fouling in a microfluidic architecture mimicking a porous microfiltration membrane. We pay special attention to how particles can overcome energy barriers leading to adsorption and desorption with each other and with the external and internal membrane surface. Interparticle interaction leads to a transition from the secondary to the primary minimum of the DLVO potential. Adsorbed particles can show re-entrainment or they can glide downstream. Since particles mainly re-suspend as clusters, the inner pore geometry significantly affects the fouling behavior. The findings allow a basic understanding of microscopic fouling events during colloidal filtration. The methodology enables future systematic studies on the interplay of hydrodynamic conditions and surface energy contributions represented by potentials for soft and patchy colloids.

*Keywords:* Fouling, Colloidal interactions, Adsorption and resuspension, CFD-DEM simulations

## 1. Introduction

Colloidal fouling is challenging in a broad range of processes such as membrane filtration, microfluidics, 3-D printing and liquid chromatography [1, 2]. In these applications fouling profoundly limit the performance and can even cause a complete failure of the process. In membrane filtration, colloidal fouling occurs inside and in front of the membrane. Colloids may be of different nature such as viruses [3], solid nanoparticles[4], and soft gel-like colloids [5]. The observation of physcial microscopic events have only become recently available through the application of sophisticated methods combining microfabrication[6], microfluidics [7, 8] and confocal laser optical analysis [9].

Despite it's importance, physical understanding of fouling is limited due to complex surface and hydrodynamic interactions on nanometer and micrometer scales. The balance between these interactions determines the fouling probability and it would be highly desirable to have a simulation methodology at hand which could resolve colloid/membrane/hydrodynamics interaction at the different scales to answer the question to what extent the geometry of porosity affects clogging under a concomittant change in flow field .

*manuscripts.cvt@avt.rwth-aachen.de
*Email address:* manuscripts.cvt@avt.rwth-aachen.de (M. Wessling)

## 2. Background

*2.1. Experimental observations*

To gain a deeper insight into the underlying events during fouling, transparent devices are commonly used to directly observe fouling behavior on a single pore level. Research focuses on the influence of three key factors, which dictate colloidal fouling: the membrane properties, the feed water composition and the hydrodynamic conditions [10]. Geometrical parameters of the membrane such as the pore size distribution and the pore shape strongly influence the fouling behavior [2, 11]. Further, looking into the mechanism on a nanoscopic level interfacial interactions between membrane and foulant play a decisive role. These surface interactions are often described by the DLVO theory, which is named after Derjaguin and Landau and Verwey and Overbeek [12–14]. According to the DLVO theory the surface interactions result from van-der-Waals and electrostatic double layer interactions. In addition to the classic DLVO theory, research showed the significant role of hydrophobic/ hydrophilic interactions on the membrane fouling [15, 16]. The surface interactions strongly depend on the ionic strength of the solution: the potential of the repulsive electrostatic double layer decreases with increasing ionic strength of the solution. Hence, an increase in ionic strength leads to a faster clogging [1, 17]. However, Sendekie et al. [17] demonstrate that at high ionic strength the clogs can become more fragile which result in re-entrainment of particle agglomerates at high flow rates. Sendekie et al. explain the fragility of clogs at higher ionic strength by diffusion-limited cluster aggregation (DLCA), which possess loose structure with low fractal dimension compared to reaction limited cluster aggregation (RLCA) formed at low ionic strength [17, 18]. Despite the variety of research and measurement techniques, fundamental fouling processes are difficult to display in experimental work due to limitations of the spatial and temporal resolution. Numerical simulations have slowly evolved over the past 20 years to overcome limitations of resolution and are used to explain mechanism of the fouling process [19–24].

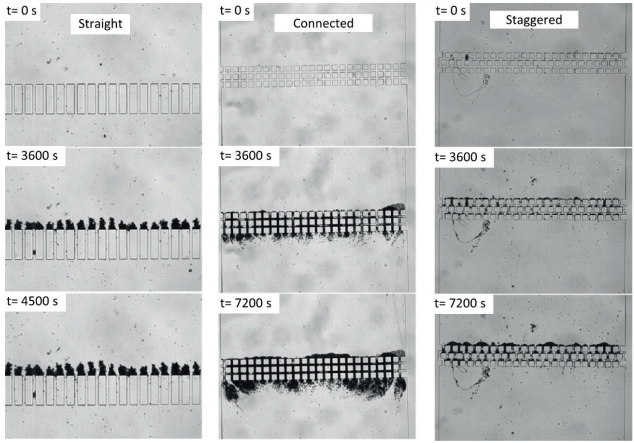

Figure 1: Experimental results showing the influence of the geometry of the micro-channel on the clogging process of 5 µm sized polystyrene particles, reprinted with permission from Bacchin et al. [25]. Copyright 2014 Springer Verlag.

### 2.2. Simulation approaches

A variety of numerical methods were used to investigate the fouling process. These range from Euler-Euler, Euler-Lagrange, fully Lagrange as well as statistical modelling approaches. Ando et al. [22] performed numerical simulations which focus on the effects of the ratio of pore and particle sized. The results show, that for higher particle diameters, relative to the pore diameter, particles form a cake layer directly on the membrane surface, while for smaller particle-pore diameter ratios the fouling first occurs in the pores and then grows to form a cake layer on the surface [22]. The effect of different particle concentration and DLVO forces on the clogging observed at pore entries has been investigated by Agbangla et al. [23]. They investigate the fouling of the entrance region of a single pore at different particle concentrations and varying repulsive forces. Without repulsive forces, cluster-dendrites form which lead to a complete blockage of the pore. When introducing repulsive forces, a minimum volume flow has to be exceeded before clogging may occur. Moreover, they observed that the particle concentration has an important effect on the development of fouling. For volumetric fractions of 5 %, fouling occurs through successive deposition, while for volumetric fractions of 20 % sudden bridging formations can appear [23].

### 2.3. Simulations and Experiments

Bacchin et al. [25] combined experiments and simulations in a microfluidic chip to investigate the influence of tortuosity and connectivity on the fouling. For this study, three different geometries were realized, which are shown in Figure 1.

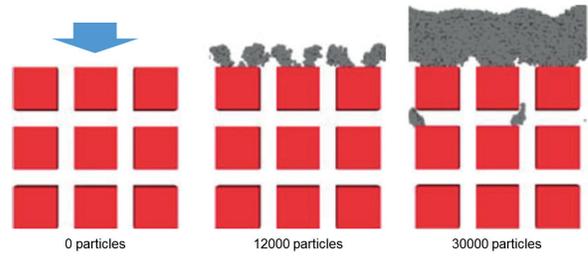

Figure 2: Simulation results of the clogging of the connected channel performed by Bacchin et al. [25], adopted and redrawn with permission from Bacchin et al. [25]. Copyright 2014 Springer Verlag.

Such structures mimic the complex geometry of membranes with three defined regular geometries referred to as the straight parallel microchannel, connected square pillars and staggered square pillars. In these microfluidic architectures, fouling experiments with 5 µm charge stabilized polystyrene particles were performed. The fouling varied significantly: while the straight channel geometry developed almost no internal fouling, the aligned square pillars are almost filled entirely with polystyrene particles, at the end of the experiment. Also the surface fouling exhibits great differences for the different geometries. In addition to the experiments, Bacchin et al. [25] compared their experimental results with simulations. Figure 2 shows Bacchin's simulation results of the clogging of the connected geometry.

Some discrepancies between the simulation and experiments exist, especially with regard to the fouling formation and the accumulation kinetics. Some important assumptions in the simulation study applied by Bacchin et al. [25] may be responsible for these differences between experiments and simulation. The used model neglects (a) particle-particle short range interactions, (b) re-suspension of adhered particles, as well as (c) the repulsive part of the DLVO forces. Another rigorous assumption dictates the walls parallel to the flow direction to be non-adherent. Due to these simplifications, the particle attachment mainly occurred artificially on the upstream side of the pillars and resembles to a certain extend the findings in [20]. This rejection in front of the membrane contradicts their experimental findings, where clogging mainly occurs on the inner structure or even at the backside.

### 3. Numerical method

We present a simulation model, which overcomes some of the mentioned limitations such as the re-suspension of particles and consideration repulsive double layer interaction, whereby we are able to demonstrate effects of the secondary minimum. The aim is to identify the events occurring during the filtration of colloidal particles which are significantly smaller than the pore size. Special attention is given on how particles overcome repulsive barriers to adsorb onto the surface and which role particle clusters can play in the adsorption and re-suspension dynamic.

The open source software CFDEM © is used to investigate the interaction of flow through a porous membrane mimic and the adsorptive behavior of colloids with each other and the membranes inner and outer surface. CFDEM © combines computational fluid dynamics (OpenFOAM ©) with a discrete element method (LIGGGHTS ©) to simulate particle motion inside a fluid flow [26, 27]. Thereby, a four way coupling algorithm determines interactions between particle-particle, fluid-particle wall-particle and wall-fluid.

*3.1. CFD-DEM approach*

In the CFD-DEM approach, the equation of continuity and the volume-averaged NavierStokes equation determine the motion of an incompressible fluid phase in the presence of a particulate phase [26].

$$\frac{\partial \alpha_f}{\partial t} + \nabla \cdot (\alpha_f u_f) = 0$$
$$\frac{\partial \alpha_f u_f}{\partial t} + \nabla \cdot (\alpha_f u_f u_f) = -\alpha_f \nabla \frac{p}{\rho_f} + \nabla \cdot \tau - R_{pf} \quad (1)$$

$\alpha_f$ and $u_f$ are the void fraction and the velocity of the fluid, respectively. $\tau$, $p$ and $\rho_f$ represent the stress tensor, the pressure and the density of the fluid. $R_{pf}$ determines the momentum exchange between the fluid and particulate phase and is calculated by the following expression [26]:

$$R_{pf} = K_{pf} \cdot (u_f - \langle u_p \rangle) \quad (2)$$

where

$$K_{pf} = \frac{\sum_i F_d}{V_{cell} \cdot |u_f - \langle u_p \rangle|} \quad (3)$$

$u_p$, $F_d$ and $V_{cell}$ represent the cell-based ensemble averaged particle velocity, the drag force and the volume of the mesh cell. To determine the momentum exchange coefficient $K_{pf}$, the drag correlation based on Lattice-Boltzmann method proposed by Koch and Hill is used [28].

The particulate phase is determined by a Lagrangian approach. Each particle trajectory is explicitly solved by the following force balance [26]:

$$m\ddot{x}_i = F_n + F_t + F_b + F_f \quad (4)$$

where $m$ and $\ddot{x}_i$ are the mass and the acceleration of the particle, respectively. The motion of each particle depends on contact forces ($F_n, F_t$), body forces ($F_b$) and forces arising due to the surrounding fluid phase ($F_f$). Gravity effects, Basset, Saffman and Magnus forces are small for this case and are therefore neglected. The Hertz contact model is used in which the contact forces depend on the length of overlap between two surfaces:

$$F_n = k_n \delta_n - \gamma_n v_n$$
$$F_t = k_t \delta_t - \gamma_t v_t \quad (5)$$

$\delta n$ and $\delta t$ denote the normal and tangential overlap of two surfaces and the normal and tangential relative velocity is termed $v_n$ and $v_t$. The constants $k_n$ and $k_t$ are the elastic constants and $\gamma_n$ and $\gamma_t$ the viscoelastic damping constants. The constants $k$ and $\gamma$ are material properties depending on the Young modulus E, the Poisson ratio $\xi$ and the coefficient of restitution $e$ [29]. Beside the hertz model, rolling friction is considered, which adds an additional torque $M_{rf}$ contribution according to the following equation:

$$M_{rf} = C_{rf} k_n \delta n r \frac{\omega_{rel,shear}}{|\omega_{rel,shear}|} \quad (6)$$

$C_{rf}$ defines the coefficient of rolling friction and $\omega_{rel,shear}$ is the projection of the relative angular velocity into the shear plane. Besides contact forces, electrostatic forces strongly influence the clogging process which are considered by the widely known DLVO theory. The DLVO theory comprises the attractive van der Waals and the repulsive electrostatic double layer potential. For particle-particle (p-p) and particle-wall (p-w) interactions both potentials are calculated as follows [30]:

Van der Waals potential:

$$E_{VDW,p-p} = -\frac{A}{6} \left( \frac{2r^2}{D(D+4r)} + \frac{2r^2}{D(D+4r)+4r^2} + \ln \frac{D(D+4r)}{D(D+4r)+4r^2} \right) \quad (7)$$

$$E_{VDW,p-w} = -\frac{A}{6} \left( \frac{2r(D+r)}{D(D+2r)} + \frac{D(D+2r)(\ln D - \ln D + 2r)}{D(D+2r)} \right) \quad (8)$$

Electrostatic double layer potential:

$$E_{EDL,p-p} = \left( \frac{r_i r_j}{r_i + r_j} \right) Z e^{-\kappa D}$$
$$E_{EDL,p-w} = r Z e^{-\kappa D} \quad (9)$$

$D$ and $r$ term the distance between particle and the wall and the particles radius, respectively. $A$ signs the Hamaker constant. The interaction constant $Z$ depends on the surface potential of particle and wall and $\kappa$ is the Debye length. Since the reciprocal function of the van der Waals potential approaches infinity at direct contact a constant force is chosen below a separation distance of 0.4 nm [31]. For a detailed discussion about the calculation of the DLVO potentials, we refer to Israelachvili and Elimelch [30, 32].

Figure 3 shows the DLVO interaction potentials over distance for 5 μm charge stabilized polystyrene particles used in the simulation among themselves and with the PDMS wall in a 100 mM NaCl solution. Based on the potentials particles can accumulate at two stable positions. The first position lies at direct contact with the surface called primary minimum, the second position is located 3 − 4 nm away from the surface termed secondary minimum. Since the attractive potential in the primary minimum are significantly higher compared to the one in the secondary minimum, particles attach stronger in the primary minimum. The characteristic of the primary and secondary

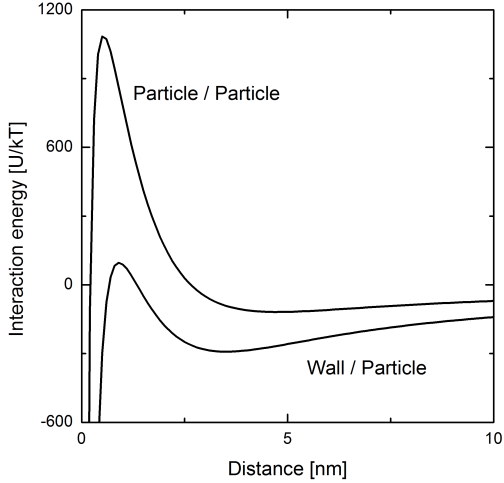

Figure 3: Computed DLVO energy profiles of particle/ particle (charge stabilized polystyrene) and particle/ wall (charge stabilized polystyrene/ PDMS) interaction in a 100 mM NaCl solution

| Parameter | Symbol | Value | Unit |
|---|---|---|---|
| Flow velocity | $u$ | 4.5 | mm s$^{-1}$ |
| Particle radius | $a$ | 5 | μm |
| Young modulus | $E$ | $3.6 \times 10^6$ | Pa |
| Poisson ratio | $\tilde{\nu}$ | 0.45 | - |
| Coefficient of restitution | $e$ | 0.95 | - |
| Friction coefficient | $\mu_f$ | 0.5 | - |
| Rolling friction coefficient | $\mu_r$ | 0.5 | - |
| Intermolecular distance | $D_{\text{int}}$ | 0.4 | nm |
| Particle density | $\rho_P$ | 1000 | kg m$^{-3}$ |
| Particle concentration | $c_P$ | 5 | vol % |
| Salt concentration | $c_{\text{KCl}}$ | 0.1 | mol L$^{-1}$ |
| Zeta potential particle | $\zeta_{PS}$ | $-37$ | mV |
| Zeta potential wall | $\zeta_{PDMS}$ | $-23$ | mV |
| Hamaker constant | $A$ | $1.4 \times 10^{-20}$ | J |
| DEM-time step | $\Delta t_{\text{DEM}}$ | $1.0 \times 10^{-8}$ | s |
| CFD-time step | $\Delta t_{\text{CFD}}$ | $1.0 \times 10^{-6}$ | s |

Table 1: Parameters applied in the simulation

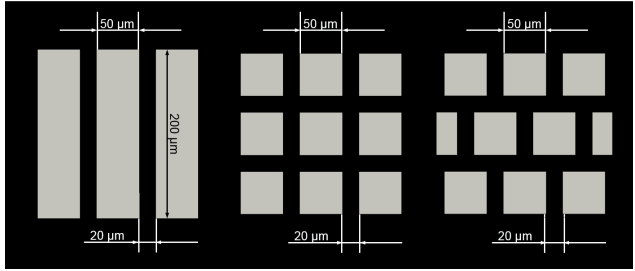

Figure 4: The straight parallel microchannels, the connected microchannels and the staggered square pillars from left to right

minimum influences the adsorption process decisively as shown by Kuznar [33].

### 3.2. Simulation method and conditions

In the following the simulation conditions, such as geometry, mesh, process conditions and simulation settings are presented. We intend to resemble the conditions of the experiments and the simulations performed by Bacchin et al. [25]. Similar to the work of Bacchin et al. [25] three types of geometries are chosen. These are referred to as the straight, the connected and the staggered geometry (see Figure 4). 5 μm sized polystyrene particles are filtered from top to bottom through the devices. The properties chosen for the PDMS walls and the particles are listed in Table 1. As in the experiments performed by Bacchin et al. [25], a constant mean inlet velocity was chosen to be 4.5 mm s$^{-1}$. For further information we refer to Bacchin et al. [25].

The computational effort strongly depends on the DEM - simulation. Therefore, the choice of the DEM - timestep is crucial to gain simulation results in a reasonable time frame. The DEM - timestep complies with the Rayleigh time, expressed by the following equation [34]:

$$t_{Coll} = 2.943 \left( \frac{5\sqrt{2}\pi\rho_P \left(1 - \nu_{Poisson}^2\right)}{4E} \right)^{\frac{2}{5}} \frac{r}{u_{Feed}} \quad (10)$$

It is good practice to set the DEM-time step to 10-30 % of the Rayleigh time and set the coupling interval with CFD to 100 DEM-time steps [35]. To reduce the computational time the Youngs modulus was chosen 3 orders of magnitude smaller than the literature value to reduce the DEM-time step. Test simulation indicate negligible differences result due to changing the Young's modulus. To further reduce the computational time, the simulations were performed with a particle concentration of 5 vol % which is 10 to 50 times higher compared to experimental work of Bacchin et al. [25]. Similar to the work of Bacchin et al. [25], simulations have been performed for 30.000 particles over a filtration time of 0.9 s. In contrast to Bacchin et al. [25], where particles are inserted batch wise, particles constantly enter the simulation domain in our model.

The DEM and the CFD are coupled with an unresolved divided volume fraction method. The particle volume is split into distributed marker points, which apportion the particle's volume to the covered mesh elements [26]. Due to this coupling of CFD and DEM, the mesh size fineness depends on the particle diameter leading to a relative coarse mesh (50.000 to 100.000 mesh elements). A possible way to improve the accuracy of the CFD and DEM coupling is to apply a resolved coupling method [26]. However, the simulation could not be performed with this method because of the computational effort.

## 4. Results & Discussion

*Macroscopic fouling and clogging*

Simulation results are performed which show the formation of colloidal fouling in three microfluidic topologies. To compare and validate the results, the geometries are based on the experimental and numerical work of Bacchin et al. [25]. Figure 5 shows the results of the fouling formation over time in the three topologies. In this Figure, particles with a velocity lower than $800\,\mu\mathrm{m\,s^{-1}}$ are displayed which keeps the focus on attached particles. The supplementary materials provides videos to show the dynamic behavior of the clogging process in the three geometries.

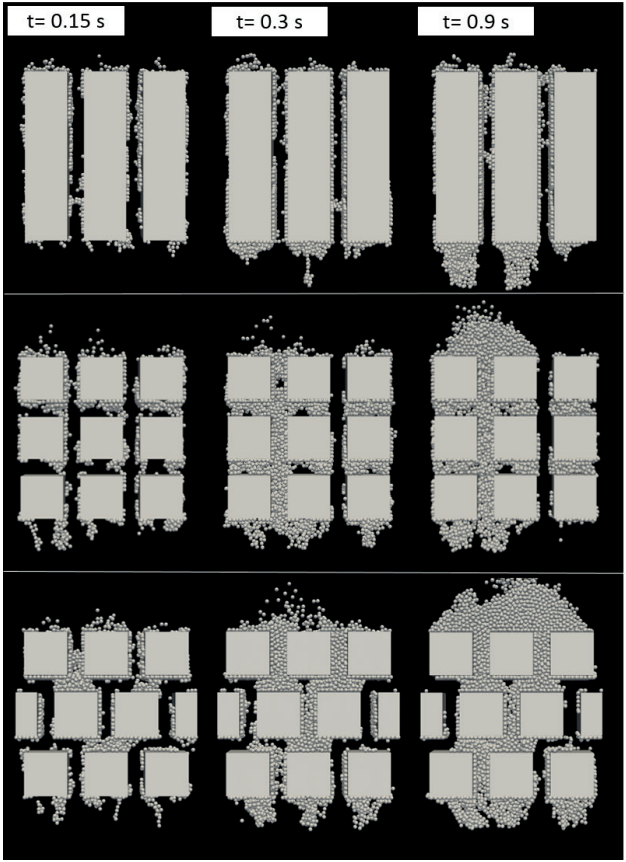

Figure 5: Simulation of the clogging development in the staggered, connected and straight geometries.

In the straight geometry, particles adsorbed on the inner surface building up a monolayer as time progresses. Also, some particle agglomerates can be observed. Interestingly, particle accumulated at the downstream zone of the pillars as well. But no complete blockage of a channel developed during the filtration. Figure 1 shows that during the experiments of the straight geometry most internal channel do not get blocked with progressing time. Instead, dendrites are formed on the inlet side of the channels. That differs from our simulation results in which most particles form dendrites on the downstream side which is discussed in Section 4.3. Furthermore, the kinetics of the fouling process between experiments and simulation are not comparable. In our simulation the fouling progresses much faster compared to the experimental results. This behavior may be explained by the assumption of smooth surfaces, neglecting of sterical interactions, acid-base interactions and the resolution of calculation of the fluid flow.

In the connected square pillars, the particle accumulate inside the inner structure which leads to a complete blockage of the left channel in course of the simulations (see Figure 5). As time proceeds, a filter cake builds up. This behavior particularly highlights the growth of a clog from the inner structure to a cake layer. Due constant volume flow condition and due to the clogging of one channel the flow velocity in the remaining channels increases, which prevents from complete clogging (see Figure 6a). Due to clogging of the channels, the pressure drop increases over time until it follows an asymptotic course (see Figure 6b). The pressure drop reaches asymptotic behavior when the center-left channel is completely blocked and the majority of the flow goes to the right channel. Pressure fluctuations are observed due to adsorption and desorption phenomena as well due to particles enter and leave the simulation domain. Therefore, the pressure drop in Figure 6b displays an averaged pressure drop over 1 ms. Similar behavior is observed in the staggered structure, in which particle adhesion starts mainly in the inner structure. Due to the cluster formations in the channels, the middle segment clogged with time and a filter cake develops. The side channels remain unclogged due to an increased flow velocity. The results of the fouling formation in the geometries with an inner structure the experiments and our simulation show good agreement. As in the simulation, particle adhered in the inner structure of the device during the experiments (see Figure 1). These adhered particles enhance further particle deposition and lead to the blockage of flow channels and a cake formation. In addition, particle deposit in the downstream zone in our simulation as well as in the aligned channel in the experiments. Bacchin et al. interpreted this downstream deposition with particle clusters which detach from the surface and settle in time. Because of the short simulation times and the little density difference between colloid and solvent, settling of particles cannot only be responsible for the deposition on the downstream side. Our model includes particle detachment as well as particle gliding which we will show later to be important microscopic events governing observations such as downstream aggregation. 4.3.

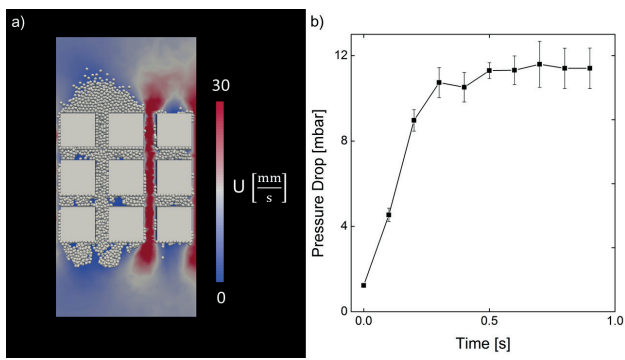

Figure 6: a) Demonstration of the flow profile through the connected channel at 0.9s. b) Development of the pressure drop as a function of time in the connected channel.

## 4.1. Transition from secondary to primary minimum

The simulations give access to all particle trajectories in time and space and enable us to investigate different microscopic events which are responsible for the clogging of the channel. One adsorption mechanism during our simulation is demonstrated in Figure 7. It exemplary demonstrates the adhesion process of a single particle (marked in blue) due to the interaction with another particle (marked in red).

At the beginning of the adsorption process of the blue marked particle, both particles are located in the bulk phase and their trajectory of movement point towards the surface of a pillar. The blue particle adsorbs into the secondary minimum with the pillar, but is not able to overcome the repulsive barrier to get into the primary minimum. The trajectories of the red particle is directed towards the blue particle leading to interaction between both particles. Due to the collision of both particles, the blue particle is pushed towards the surface of the pillar. The blue particle overcomes the repulsive barrier of the particle-wall potential and reaches the primary minimum. Due to the higher repulsive barrier of the particle-particle potential both particle do not agglomerate. The red colored particle re-suspends in the solution again due to the fluid flow.

This adsorption process is even more pronounced at the side walls of the pillars (8). Due to high kinetic energy of the particles in the channels, the particle collision leads to a transition from secondary to primary minimum on the wall as well as an agglomeration of both particles. This clustering deposition is experimentally observed during a filtration experiment by Zamani et al. [16]. They showed that a glass particle sticking to a membrane surfaces caught a particle from the bulk, which leads to a deposition on the membrane surface.

In summary, particles overcome the repulsive barrier due to particle particle interactions and the attached particles transits into the primary minimum.

## 4.2. Particle adsorption on the edges

In contrast to the previous section which deals with the adsorption of particles due to particle-particle interactions, this section shows the adsorption of single particles into the primary minimum. To reach the primary minimum the drag force needs to be sufficient to overcome the repulsive barrier. Particularly at the entrance of the pillars the particles are sufficiently accelerated to reach the necessary potential. Therefore, particle deposit on the edge of the pillars as shown in Figure 9.

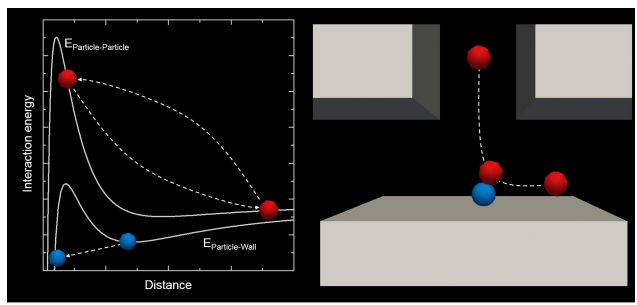

Figure 7: Particle-particle interaction leading to overcome of the repulsive barrier of the wall-particle potentials (blue colored particle). Due to the higher repulsive barrier of the particle/ particle potential both particles do not agglomerate. Therefore, the fluid flow entrains the red colored particle.

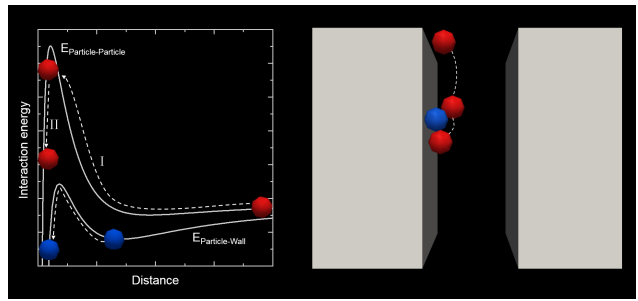

Figure 8: Particle-particle interaction leading to overcome of the repulsive barrier of the wall-particle potentials as well as of the particle-particle potentials, whereby an agglomerate is formed on the surface of the pillar.

The adsorption of particles at the corners of the constriction was also observed in several experimental studies [17, 21, 36, 37]. This mechanism of adsorption is relevant to comprehend the bridge formation of particles leading to blockage of the channel [8].

Both phenomena, particle adsorption on the edges as well as adsorption due to collision, play a decisive role in the clogging dynamic. During our simulation the adsorption due to collision was the dominant mechanism because of the high collision frequency at high particle concentration. Further parameters such as the pore size to particle diameter and the velocity influence the behavior of the clogging dynamic.

## 4.3. Dynamics of the clogging process

A variety of numerical studies treat particles in contact with the surface as fixed and do not solve the equations of motion for these particles [23, 25, 38]. In contrast to these studies, we are able to show their dynamic behavior even though they might have been adsorbed initially. Now, one observes that particles glide over the surface as demonstrated in Figure 10 during the clogging process,. Due to gliding, particles can deposit in the downstream zone of the pillars. A similar particle behavior is observed experimentally by Sendikie et al. [17] (refer Figure 10). In their study, polystyrene particles glide downstream the pillars at 10 mM salt concentration leading to clogging in the downstream corner. The evolution and the velocity of gliding particles is overestimated in our model formulation however. The reason lies in the assumption of a perfectly smooth surface used in the modelling approach. Therefore, particles located in the secondary minimum do not experience friction forces leading to an unhindered movement tangential to the PDMS surface. These particles tend to move over the corners to the center of the downstream surface, which results in increased particle deposition on the downstream surface. Nonetheless, the gliding of particles downstream can be decisive in the clogging dynamic, especially in systems with an inner structure such as membranes.

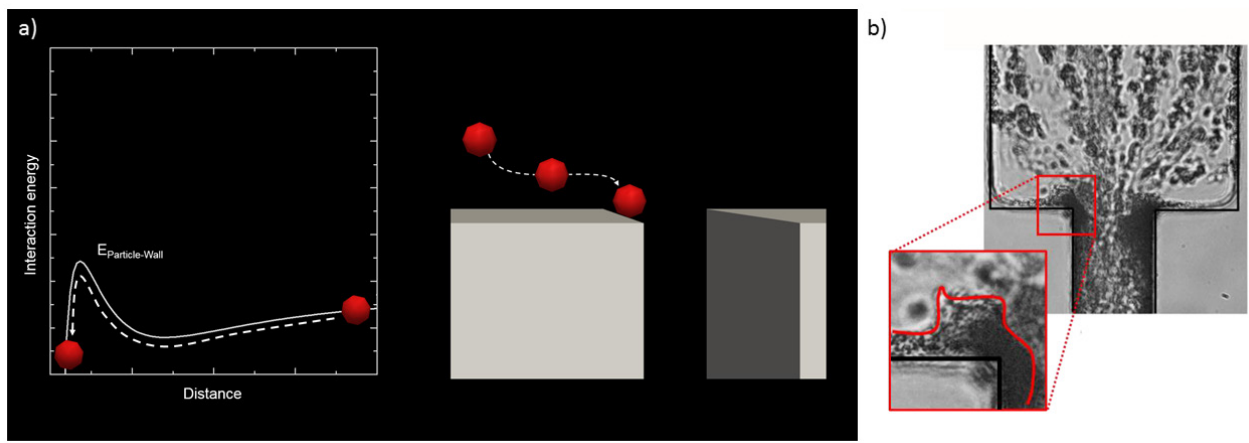

Figure 9: a) Schematic drawing of a particle adsorption on the edges of the pillars in the simulations. b) Experimental results of particle deposition on the edges, reprinted with permission from Lee et al. [36]. Copyright 2017 Elsevier

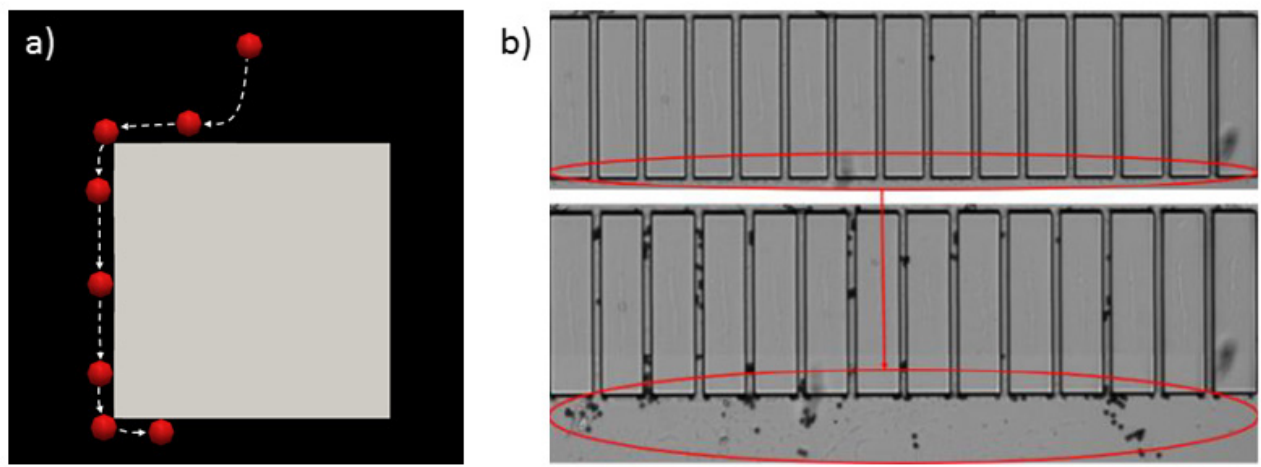

Figure 10: a) Particle glides over the surface leading to attachment on the downstream zone of the pillars during simulation. b) Particle capture on the downstream corner due to particle gliding in experimental study of Sendiecke et al. [17], reprinted with permission from [17]. Copyright 2016 American Chemical Society.

*4.4. Re-entrainment and re-attachment*

During the clogging process, resuspension phenomena are frequently observed. Particles in the first monolayer on the surface show strong attachment and nearly do not re-entrain. In contrast, interparticle bonds in a particle multilayer frequently break. The increasing velocity to the middle of the channel and the smaller energy necessary for desorption of interparticle interaction compared to particle/ wall interaction explain the breakage of particle bonds. The breakage of particle agglomerates is far more often observed than the re-entrainment of single particles. The resuspension of a particle agglomerate is demonstrated in Figure 11. Robert de Saint Vincent et al. [39] showed experimentally that polystyrene particles were swept out from a PDMS surface at high flow velocities. Sendiecke et al. [17] reported at high ionic strength (100 mM salt) that clogging of some channels occurred, which are labile and frequently break away. The agglomerates are pushed into the inner structure of the device. Due to their large volume, the agglomerates tend to re-attach again as demonstrated in Figure 11. These agglomerates serve as initiator for the complete blocking of a channel. Due to flow deflections in the inner structure the attachment of agglomerates is significantly amplified. Amongst other things, this explains why the straight channel does not show a complete blockage of a channel during the simulation in contrast to both other devices (see Figure 5). This mechanism demonstrates the importance of an inner structure on the clogging process.

## 5. Future challenges

Improving the resolution of the hydrodynamic calculation will be necessary to match length scales of hydrodynamic and particle interactions. Currently, the simulation average the resolution depth of the hydrodynamics on a micrometer scale whereas the particle surface interactions are regarded on a nanometer scale. By increasing the resolution of the fluid simulation lubrication effects can be determine more accurately. Lubrication means the "squeezing" of liquids between two surfaces leading to an additional adsorption resistance. Such effect is suggested to influence particle deposition [40]. Also, particle adhesion strongly depends on surface heterogeneities such as surface roughness and unequal surface charge distributions [41]. A mean field approach for colloidal interactions as applied in this work may not be sufficient to account for these surface non-idealities. Also, it will be important to answer the question whether the chosen DLVO potential is representative, or whether an extended DLVO potential would be more appropriate [42]. While these questions can be patiently answered with simulations, the experimental support needs to be developed as well. Hence, the reported finding can only be considered as another contribution to the puzzles of colloidal membrane fouling and requires further experimental proof.

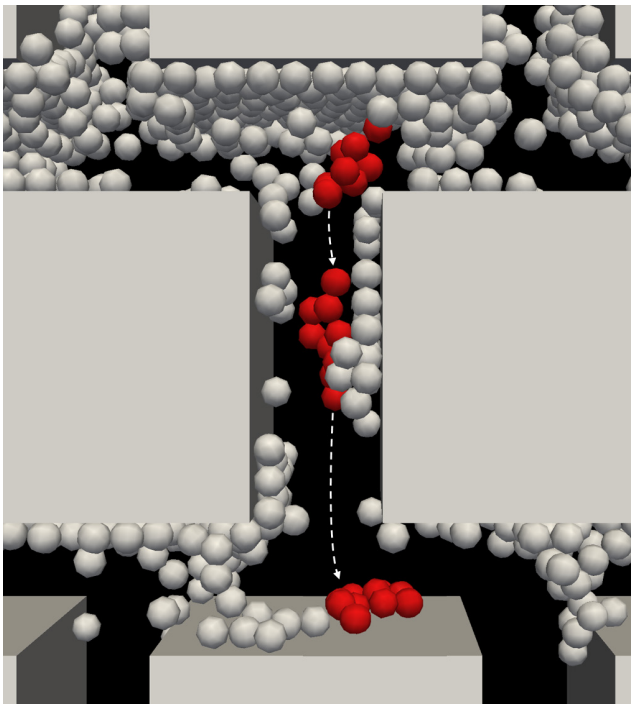

Figure 11: Re-entrainment and re-attachment of a particle cluster

## 6. Conclusion

We carried out numerical simulations of a fouling process in a microfluidic membrane mimic using an CFD-DEM approach. The simulation results show good agreement with the experimental work of Bacchin et al. and we discuss the improvements compared to their simulation [25]. The presented method identifies important microscopic events of the clogging process and compares them to experimental findings. Interparticle interactions have a strong effect on the clogging dynamics. Adsorbed particles can re-entrain from the inner membrane surface or they can glide downstream. The inner structure of the porous membrane can significantly affect the clogging process. In particular the re-entrainment of particle clusters and their reattaching to the inner surface can lead to a complete blockage of the flow channel.

The methodolgy presented allows to quantify the transition process of particles from the secondary to primary minimum adsorption during filtration. The latter is almost undetectable in an experimental study as it would require of a nanometer spatial resolution and a very high temporal resolution.

The methodology now enables to perform systematic studies comparing experiments and simulations leading to a more comprehensive understanding of the deposition phenomena in micron and sub-micron sized porous filtration media. It encourages and directs experimentalist and simulation scientist to further join forces and unravel the intricacies of colloidal deposition in porous media such as synthetic porous membranes.

## Acknowledgments

Matthias Wessling acknowledges the Alexander-von-Humboldt Foundation and the European Research Council Advanced Investigator Programm (694946) for financial support. We thank Professor Naegele (FZ Juelich) and his team for reflection and discussions.